\documentclass[%
showpacs,preprintnumbers,
 amsmath,amssymb,
 aps,
 prd,
 lengthcheck,
letterpaper
]{aastex6}

\usepackage{amsmath}
\usepackage{graphicx}
\usepackage{scrextend}
\usepackage{hyperref}
\usepackage{txfonts}
\usepackage{color}
\usepackage{multirow}

\newcommand{\noop}[1]{}

\newcommand{\be}{\begin{equation}}
\newcommand{\ee}{\end{equation}}
\newcommand{\beq}{\begin{eqnarray}}
\newcommand{\eeq}{\end{eqnarray}}

\def\nue{\mathrel{{\nu_e}}}
\def\numu{\mathrel{{\nu_\mu}}}
\def\nutau{\mathrel{{\nu_\tau}}}
\def\nux{\mathrel{{\nu_x}}}

\def\barnue{\mathrel{{\bar \nu}_e}}
\def\barnumu{\mathrel{{\bar \nu}_\mu}}
\def\barnutau{\mathrel{{\bar \nu}_\tau}}
\def\barnux{\mathrel{{\bar \nu}_x}}

\def \lta {\mathrel{\vcenter{\hbox{$<$}\nointerlineskip\hbox{$\sim$}}}}
\def \gta {\mathrel{\vcenter{\hbox{$>$}\nointerlineskip\hbox{$\sim$}}}}

\def\t13{\mathrel{{\theta_{13}}}}
\def\y12{\mathrel{{\tan^2 \theta_{12}}}}
\def\c2{\mathrel{{\chi^2 }}}

\def\msun{\mathrel{{M_\odot }}}

\newcommand{\kmpers}{\ensuremath{\rm{km}\ \rm{s}^{-1}}}
\newcommand{\maxdm}{{\ensuremath{\Delta {\mathrm M}_{\rm{max}}}}} 
\newcommand{\net}{{\tt mesa\_{204}.net}} 

\newcommand{\n}{neutrino}
\newcommand{\ns}{neutrinos}
\newcommand{\ps}{presupernova}
\newcommand{\sn}{supernova}

\newcommand{\oss}{oscillations}

\newcommand{\bp}{$\beta$p}

\begin{document}

\title{Neutrinos from beta processes in a presupernova: probing the isotopic evolution of a massive star   }

\author{Kelly M.\ Patton}
\email{kmpatton@uw.edu}
\affiliation{Institute for Nuclear Theory, University of Washington, Seattle, WA 98195 USA}

\author{Cecilia Lunardini}
\email{cecilia.lunardini@asu.edu}
\affiliation{Department of Physics, Arizona State University, Tempe, AZ 85287-1504 USA}

\author{Robert J.\ Farmer}
\email{r.j.farmer@uva.nl}
\affiliation{Anton Pannenkoek Institute for Astronomy, University of Amsterdam, NL-1090 GE Amsterdam, the Netherlands}
\affiliation{School of Earth and Space Exploration, Arizona State University, Tempe, AZ 85287-1504 USA}

\author{F.~ X.~Timmes}
\email{ftimmes@asu.edu}
\affiliation{School of Earth and Space Exploration, Arizona State University, Tempe, AZ 85287-1504 USA}
\affiliation{JINA, Joint Institute for Nuclear Astrophysics, USA}

\begin{abstract}
We present a new calculation of the \n\ flux received at Earth from a massive star in the $\sim 24$ hours of evolution prior to its explosion as a supernova (presupernova).  Using the stellar evolution code MESA, the \n\ emissivity in each flavor is calculated at many radial zones and time steps.  In addition to thermal processes, \n\ production via beta processes is modeled in detail, using a network of 204 isotopes.  We find that the total produced $\nue$ flux has a high energy spectrum tail, at $E \gta 3 - 4$ MeV, which is mostly due to decay and electron capture on isotopes with $A = 50 - 60$. In a tentative window of observability of $E \gta 0.5$ MeV and $t < 2$ hours pre-collapse, the contribution of beta processes to the $\nue$ flux is at the level of  $\sim90$\% . For a star at $D=1$ kpc distance, a 17 kt liquid scintillator detector would typically observe several tens of events from a presupernova, of which up to $\sim 30\%$ due to beta processes.  These processes dominate  the signal at a liquid argon detector, thus greatly enhancing its sensitivity to a presupernova. 

\end{abstract}

\date{\today}

\maketitle


\section{Introduction}
\label{sec:intro}

The advanced evolution of massive stars -- that culminates in their collapse, and possible explosion as supernovae -- has been observed so far only in the electromagnetic band. 
Completely different messengers, the neutrinos, dominate the star's energy loss from the core carbon burning phase onward, and, with their fast diffusion time scale, they set the very rapid pace (from months to hours) of the latest stages of nuclear fusion ({\it presupernova}). These neutrinos have never been detected; their observation in the future would offer a  
  unique and direct probe of the physical processes that lead to stellar core collapse.   

  In a star's interior, \ns\ are produced via a number of thermal processes -- mostly pair production --  and via $\beta$-processes, i.e., electron/positron captures on nuclei and nuclear decay. The \n\ flux from thermal  processes mainly depends on the thermodynamic conditions in the core. The neutrino flux from $\beta$ reactions have a stronger dependence on the isotopic composition, and thus on the complex network of nuclear reactions that take place in the star.  In this respect, the two classes of production, thermal and $\beta$, carry complementary information.

  At this time, the study of the thermal \n\ flux from a \ps\ star is fairly mature. Exploratory studies in 2003-2010   \citep{Odrzywolek:2003vn, Odrzywolek:2004em, Kutschera:2009ff, Odrzywolek:2010zz}  showed that they can be detected in  the largest neutrino detectors for a star at a distance $D \lta 1$ kpc. Later, detailed descriptions of the thermal processes \citep{Ratkovic:2003td, Odrzywolek:2007xp, Dutta:2003ny, Misiaszek:2005ax} have been applied to state of the art numerical simulations of stellar evolution, to obtain the time dependent \ps\ \n\ flux expected at Earth \citep{Kato:2015faa,Yoshida:2016imf}. The potential of \ps\ \ns\ as an early warning of an imminent nearby \sn\ was emphasized \citep{Yoshida:2016imf}.

For the \ns\ from $\beta$ processes (henceforth \bp), the status is very different. Dedicated studies have developed much more slowly, as it was  recognized early on \citep{Odrzywolek:2009wa,Odrzywolek:2010zz} that they required a complex numerical study of realistic stellar models with large nuclear networks.   

In a recent publication \citep{Patton:2015sqt}, we have approached the challenge of modeling the \bp\ in detail -- in addition to the thermal processes --  for a realistic, time evolving star simulated with the MESA software instrument \citep{Paxton:2010ji, Paxton:2013pj, Paxton:2015jva}. The 204 isotope nuclear network of MESA, fully coupled to the hydrodynamics during the entire calculation, made it possible to obtain, for the first time, consistent and detailed emissivities and energy spectra for the $\beta$ \ns, at sample points inside the star at selected times pre-collapse. It was found that \bp\  contribute strongly to the total \n\ emissivities, and even dominate at late times and in the energy window relevant for detection ($E \gta 2$ MeV or so).
Using an independent numerical simulation, with a combination of nuclear network and arguments of statistical equilibrium, \citet{Kato:2017ehj} reached similar conclusions, and calculated the  rates of events expected in \n\ detectors as a function of time as well as total numbers of events.

In this paper, we further extend the study of \ps\ \ns, with emphasis on a realistic, consistent description of the flux from \bp. For two progenitor stars, evolved with MESA, the time-dependent \n\ emissivities for different production processes are integrated over the volume of emission, so to obtain 
the neutrino luminosities and energy spectra expected at Earth for each \n\ flavor.
For several time steps leading to the collapse, the isotopes that dominate the \bp\ emission, for both neutrinos and antineutrinos, are identified. 
We discuss the prospects of detectability, and how they depend on the distance to the star, ranging from the nearby Betelgeuse to progenitors as far as the horizon of detectability, beyond which no observable signal is expected. In the discussion, the main guaranteed detector backgrounds are taken into account.

The paper is structured as follows.  In Sec. \ref{sec:theory}, a concise summary of our simulation is given. Sec. \ref{sec:fluxes} gives the results for the \n\ flux and energy spectrum produced in a \ps\ star as a function of the time pre-collapse.  Sec. \ref{sec:atearth} shows the expected \n\ flux at Earth, with a brief discussion of \oss\ effects and detectability. A discussion follows in Sec. \ref{sec:discussion}. 

\section{Neutrino production and stellar evolution }
\label{sec:theory}

We simulate the evolution of two stars of initial masses $M=15, 30~\msun$ (with $\msun$ the mass of the Sun), from the pre-main sequence phase
to core collapse, using MESA r7624 \citep{Paxton:2010ji, Paxton:2013pj, Paxton:2015jva}, for which inlists and stellar models used are publicly available \footnote{\url{http://mesastar.org}}.
The MESA runs used here are the same as those in \citet{Farmer:2016aaa}, where technical details can be found.   Each star is modeled as a single, non-rotating, non-mass losing, solar metallicity object. The calculation stops at the onset of core collapse, which is defined as the time when any part of the star exceeds an infall velocity of 1000 \kmpers. We also set the maximum mass of a grid zone to be $\maxdm=0.1\msun$. 
The simulations employ a large, in-situ, nuclear reaction network,  \net, consisting of 204 isotopes up to $^{66}$Zn, and including all relevant reactions.
They also include effects of convective overshoot, semi-convection and thermohaline mixing on the chemical mixing inside the star.

In output, MESA gives the time- and space-profiles of the temperature $T$, matter density $\rho$, isotopic composition,  and electron fraction, $Y_e$.  These quantities are then used in a separate calculation to derive the neutrino fluxes, as outlined in our previous work \citep{Patton:2015sqt}. For brevity, here only the main elements are summarized. 

We calculate the spectra for $\nue$, $\barnue$ and $\numu,\nutau,\barnumu,\barnutau$ (collectively called $\nux$ and $\barnux$ from here on) resulting from $\beta$ processes and pair annihilation. Other thermal processes \citep{Patton:2015sqt} were found to be by far subdominant in the  late time \n\ emission from the whole star, and were neglected for simplicity.

In the calculation of spectra for the \bp, the relevant rates are taken from the nuclear tables of Fuller, Fowler and Newman (FFN) \citep{Fuller:1980zz, Fuller:1981mt, Fuller:1981mv, Fuller:1085zz}, Oda \textit{et al.} (OEA) \citep{Oda:1994} and Langanke and Martinez-Pinedo (LMP) \citep{Langanke:2001td}.  For isotopes that appear in multiple tables, the rates of LMP are given precedence, followed by OEA, then finally FFN. This order of precedence is the same as used in MESA \citep{Paxton:2010ji}.  

As described in FFN \citep{Fuller:1980zz, Fuller:1981mt, Fuller:1981mv, Fuller:1085zz}, the rate of decay from a parent nucleus in the excited state $i$ to a daughter in excited state $j$ is
\begin{equation}
\lambda_{ij} = \log{2}\frac{f_{ij}(T,\rho,\mu)}{\langle ft \rangle_{ij}}.
\end{equation}
Here, $\langle ft \rangle_{ij}$ is the comparative half-life, containing all of the nuclear structure information and the weak interaction matrix element.  The function $f_{ij}$ is the phase space of the incoming and outgoing electrons or positrons.  It uniquely determines the shape of the resulting \n\ spectrum, because the outgoing neutrinos are presumed to be free streaming with no Pauli blocking. 

Since the shape of the spectrum is entirely determined by the phase space, we can define the spectrum as $\phi = N \, f_{ij}(T,\rho,\mu)$, where $N$ is a normalization factor.  We then write the spectra for the \bp\ \ns\ for a single isotope as:
\begin{eqnarray}
\phi_{EC,PC}   &= & N \frac{E_{\nu}^{2}(E_{\nu} - Q_{ij})^{2}}{1 + \exp{((E_{\nu} - Q_{ij} - \mu_{e})/kT)}}   \Theta (E_{\nu} - Q_{ij} - m_{e})
\label{betaspec1}
\\
\phi_{\beta}  & = & N \frac{E_{\nu}^{2}(Q_{ij} - E_{\nu})^{2}}{1 + \exp{((E_{\nu} - Q_{ij} + \mu_{e})/kT)}}  \Theta (Q_{ij} - m_{e} - E_{\nu}),
\label{betaspec2}
\end{eqnarray}
where EC (PC) is for electron (positron) capture, and $\beta$ is for decay.  The chemical potential $\mu_{e}$ is defined including the rest mass such that $\mu_{e^{-}} = -\mu_{e^{+}}$.  The parameter $Q_{ij} = M_{p} - M_{d} + E_{i} - E_{j}$ is the $Q$-value for the transition, where $M_{p,d}$ is the mass of the parent (daughter) and $E_{i,j}$ is the excitation energy.   

The rates reported in the FFN, OEA, and LMP tables are actually the sum of all possible transitions, so $\lambda = \Sigma \lambda_{ij}$.   So rather than finding individual values for each $Q_{ij}$, we follow the method of \citet{Langanke:2001td} and \citet{Patton:2015sqt}, and instead find an effective $Q$-value.  We calculate the spectrum and its average energy, then adjust the $Q$-value until the average energy in the rate tables is reproduced.  Note that the tabulated average energy is a combined value for both decay and capture, therefore $Q$ is the same for both processes.  

The parameter $N$ in Eqs. (\ref{betaspec1})-(\ref{betaspec2}) is a normalization factor, defined to reproduce the tabulated rates $\lambda_i$ for isotope $i$:
\begin{eqnarray}
\lambda_{i} = \int_{0}^{\infty} \phi_{i} dE_{\nu} \,\,\,\,\,\,\, i = EC, PC, \beta^{\pm}.
\end{eqnarray}

The total spectrum of \ns\ from \bp\ (comprehensive of both capture and decay processes)  is given by the sum over all the isotopes, weighed by their abundances $X_k$: 
\begin{equation}
\left(\frac{dR_{\beta}}{dE}\right)_{\nu_{e},\bar{\nu}_{e}} = \sum_{k} X_{k} \phi_{k} \frac{\rho}{m_{p}A_{k}}.
\label{totalPhiBeta}
\end{equation}
Here $m_{p}$ is the mass of the proton, and  $A_{k}$ is the atomic number of the isotope $k$.  

For \ns\ produced via pair annihilation, the emission rate, differential in the \n\ energy, is
\begin{equation}
\left(\frac{dR}{dE}\right)_{\nu_{\alpha},\bar{\nu_{\alpha}}} = \int d^{3}{p}_{1}d^{3}{p}_{2}\left(\frac{d\sigma \varv}{dE}\right)_{\nu_{\alpha},\bar{\nu_{\alpha}}} f_{1}f_{2},
\label{dRdEpair}
\end{equation}
where $f_{i}$ is the Fermi-Dirac distribution function for the electron and positron, and
\begin{equation}
d\sigma~ \varv = \frac{1}{2\mathcal{E}_{1}} \frac{1}{2\mathcal{E}_{2}} \frac{1}{(2\pi)^{2}} \delta^{4}(P_{1} + P_{2} - Q_{1} - Q_{2}) \frac{d^{3}q_{1}}{2 E_{1}} \frac{d^{3}q_{2}}{2 E_{2}}   \langle | \mathcal{M} |^{2} \rangle~ .
\end{equation}
Here, $\varv$ is the relative velocity of the electron-positron pair, $P_{1,2} = (\mathcal{E}_{1,2}, \bf{p}_{1,2})$ is the four-momentum of the electron (positron), and $Q_{1,2} = (E_{1,2}, \bf{q}_{1,2})$ is the four-momentum of the (anti-)\n.  
The squared matrix element, as given by \citet{Misiaszek:2005ax}, is 
\begin{eqnarray}
\langle | \mathcal{M} |^{2} \rangle = 8 G_{F}^{2} \left( \left( C_{A}^{f} - C_{V}^{f}\right)^{2}\left(P_{1}\cdot Q_{1}\right) \left( P_{2}\cdot Q_{2}\right) 
+ \left( C_{A}^{f} + C_{V}^{f}\right)^{2}\left(P_{2}\cdot Q_{1} \right) \left(P_{1}\cdot Q_{2} \right)
+ m_{e}^{2} \left( C_{V}^{2} - C_{A}^{2}\right) \left(Q_{1}\cdot Q_{1}\right)\right).
\end{eqnarray}
Here, $C_{V}^{e} = 1/2 + 2\sin^{2}{\theta_{W}}$, $C_{A}^{e} = 1/2$, and $C_{V,A}^{x} = C_{V,A}^{e} - 1$.

In \citet{Patton:2015sqt}, Eq. (\ref{totalPhiBeta}) and (\ref{dRdEpair}) were used to calculate the spectra for selected times and points inside a star.  Here, we integrate over the emission region, to obtain the number luminosity -- i.e., the number of \ns\ that leave the star per unit time -- and the differential luminosity:

\begin{eqnarray}
\frac{dL^{\nu_{\alpha}}_{N}}{dE} & = & 4 \pi \int{\left(\frac{dR}{dE}\right)_{\nu_{\alpha}} r^{2} dr}, \\
L^{\nu_{\alpha}}_{N} & = & \int{\frac{dL^{\nu_{\alpha}}_{N}}{dE} dE}.
\label{luminosityEq}
\end{eqnarray}

\section{Results: time profiles and spectra}
\label{sec:fluxes}

Results  were obtained for discrete times (time-to-collapse, $\tau_{CC}$)  between the onset of core oxygen burning and the onset of core collapse.  An interval of two hours prior to collapse -- when the chance for detection is greatest -- was mapped in greater detail. Specifically, for the $15\msun$ (30 $\msun$) model, we took a total of 21 (26) time instants, of which 15 (20) in the final two hours.  All the calculated times are shown in  Fig. \ref{lumVsTime}, while a subset of seven times is  investigated in more detail in other  figures and tables. 
Calculations of numbers of events in detectors use all the calculated times within the last two hours.

\subsection{A neutrino narrative: time-evolving luminosities}
\label{subsec:lum}

\begin{figure*}
\begin{centering}
\includegraphics[width =0.85 \linewidth]{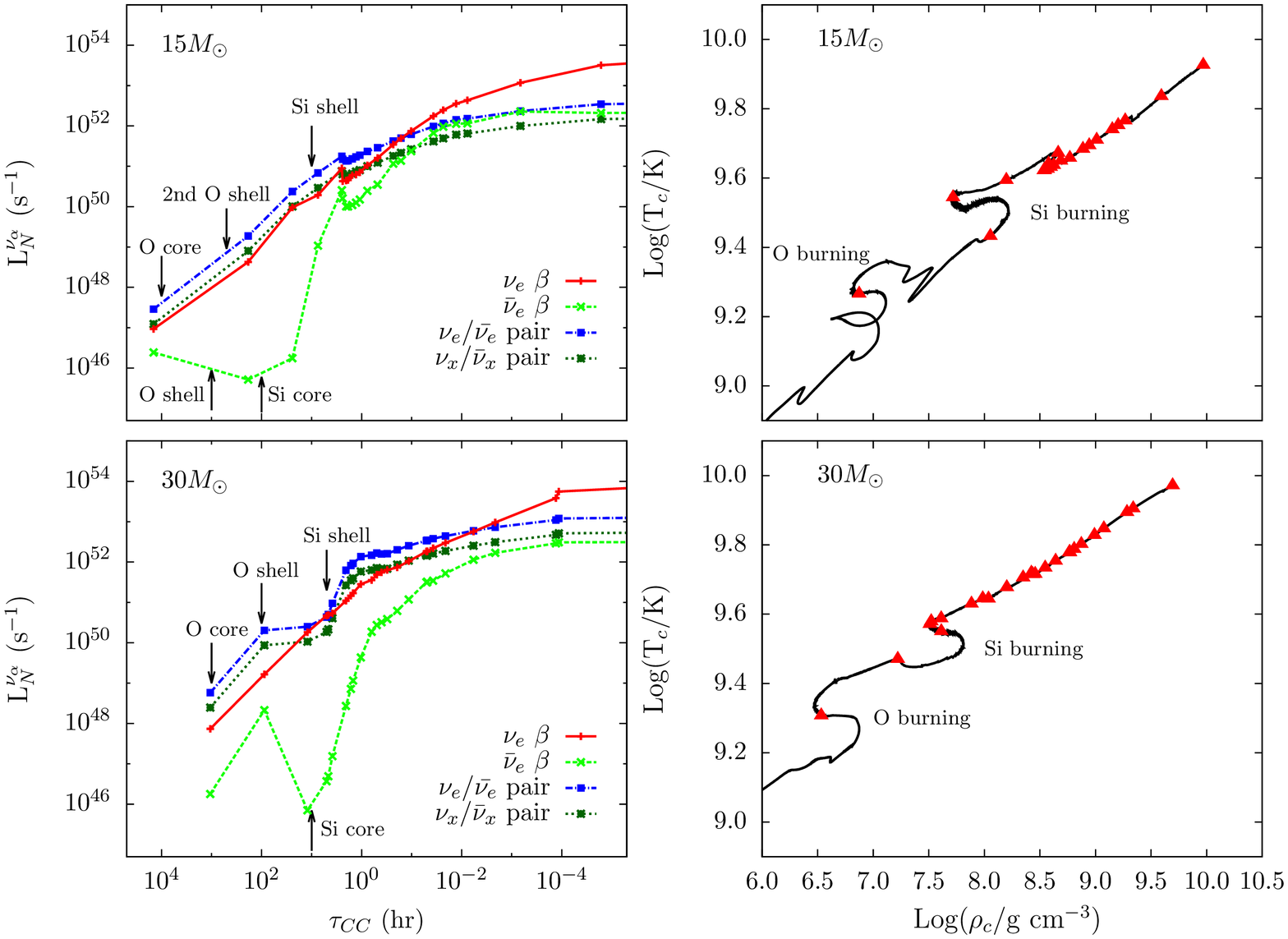}
\caption{The time evolution of the two progenitors of mass $M=15,30 \msun$. Here $\tau_{CC}$ is the time-to-collapse (in hours). In all figures, the markers correspond to the points at which the \n\ luminosities were calculated. {\it Left:} Total number luminosities for different production channels: $\nue$ from \bp,  $\barnue$ from \bp, $\nue$/$\barnue$ from pair annihilation and $\nux$/$\barnux$ from pair annihilation. The arrows indicate approximate times of ignition for the different fuels.   {\it Right:} The trajectory in the plane of central temperature and central density. } 
\label{lumVsTime}
\end{centering}
\end{figure*}

Let us examine the thermal history of the two progenitors, and how it is reflected in the \n\ luminosity. Fig. \ref{lumVsTime} shows the star's trajectory in the plane of central temperature and central density, $(T_{c},~ \rho_{c})$, as the time evolves. It also shows the evolution for the neutrino number luminosities, $L^{\nu_{\alpha}}_{N}$, for different production channels, and the approximate times of ignition of the various fuels.

From the figure, it appears that the evolution of the two stars is generally similar, the main difference being that the more massive progenitor evolves faster and is overall brighter in \ns. 
In particular, for the 15 $\msun$ (30 $\msun$) star
the burning stages for the two stars proceed as follows: at $\tau_{CC}\approx 10^4$ hrs ($\tau_{CC}\approx 10^3$ hrs),  oxygen ignition takes place in the core, and proceeds convectively until it ceases at $\tau_{CC}\approx 10^3$ hrs  ($\tau_{CC}\approx 10^2$ hrs). Then, an oxygen shell is ignited and burns until $\tau_{CC}\approx 5\times10^2$ hrs ($\tau_{CC}\approx 10$ hrs).  Eventually, silicon burning is ignited in the core and proceeds until $\tau_{CC}\approx 10$ hrs ($\tau_{CC}\approx 5$ hrs).   At that point the star transitions to
shell silicon burning, which proceeds until collapse.  Interestingly, the 15 $\msun$ star has an intermediate phase (which is absent in the more massive progenitor) before core silicon burning: a second, off center oxygen burning stage,  which lasts until $\tau_{CC}\approx 10^2$ hrs.

In Fig. \ref{lumVsTime}, we can see how the luminosity of $\nue$ from \bp\ grows faster than that of thermal processes. For the $15 \msun$ (30 $\msun$) case, it 
amounts to $\sim$30$\%$ ($\sim$10$\%$) of the contribution from pair annihilation at the onset of oxygen burning; it becomes comparable to pair annihilation  
at $\tau_{CC}\approx6$ min ($\tau_{CC}\approx7$ s), increasing to almost an order of magnitude greater ($\sim$30 times greater) at the onset of core collapse.

The luminosity of $\barnue$ from \bp\ follows a more complicated pattern, tracing more closely the phases of stellar evolution.  It drops after core oxygen burning ends, and begins increasing again after silicon core ignition. 
The total $\barnue$ emission is always dominated by  pair annihilation, although the disparity decreases as the stars approach core collapse.  At the onset of core collapse, the \bp\ contribution is approximately 40$\%$ ($\sim 20\%$) of the pair process for the 15 $\msun$ (30 $\msun$ model) model.  

A unique feature of the 15 $\msun$ model is a short sharp drop in the luminosities of all \n\ species, 
shortly after shell silicon burning begins, followed by a smooth increase.  
This peak is absent in the time profiles of the 30 $\msun$ model, for which the time profiles are smoother. This difference can be traced to differences in the core carbon burning phases of the two stars, which proceed convectively for the $M=15\msun$ case and radiatively for the $M=30\msun$ model \footnote{The dividing line between the two paths is given by the central carbon mass fraction, with critical value X($^{12}$C)$\sim$20\% \citep{1993PhR...227...65W, 1996ApJ...457..834T, 2002RvMP...74.1015W}. For the MESA inputs used here, solar metallicity models with Zero Age Main Sequence masses below $\simeq~20~\msun$ have X($^{12}$C)$\gtrsim$20\% and thus undergo convective core Carbon-burning. See, e.g., \citep{petermann_2017_aa}.}. For convective core C-burning, efficient neutrino emission decreases the entropy. This entropy loss is missing in the radiative carbon burning case, causing all subsequent burning stages to take place at higher entropy,  higher temperatures, and lower densities.  In these conditions, density gradients are smaller and extend to larger radii, thus explaining the smoother profiles of the 30 $\msun$ model.

We notice that the \n\ luminosity from pair annihilation increases more slowly in the last few hours of evolution.  This can be understood considering that the emissivity for pair annihilation is nearly independent of the density for fixed temperature \citep{Itoh:1996im}, and therefore directly reflects the moderate increase of the temperature (Fig. \ref{lumVsTime}, right panes) over hour-long periods.

Generally, the patterns found here are consistent with those  in the recent work by \citet{Kato:2017ehj}. The main difference is in the $\barnue$  luminosity from \bp, which in our work is always subdominant, while in  Kato \textit{et al.} it dominates over pair annihilation starting at $\tau_{CC}\sim 0.5$ hrs. This discrepancy could be due to the nuclear networks used: in our work, the network \net\ is evolved self-consistently within MESA to obtain mass fractions, and tabulated  \bp\ rates  from FFN, ODA, and LMP are used (see Sec. \ref{sec:theory} and \citep{Patton:2015sqt}). Instead, Kato \textit{et al.} calculate mass fractions using nuclear statistical equilibrium, and incorporate many neutron rich isotopes, with rates taken from tables by Tachibana and others \citep{Tachibana:1995, Yoshida:2000, Tachibana:2000, Koura:2003, Koura:2005}, which they adapted to the stellar environment of interest (the original tables are for terrestrial conditions).

\subsection{Neutrino spectra: isotopic contributions}
\label{subsec:spec}

\begin{figure*}
\begin{centering}
\includegraphics[width =0.8 \linewidth]{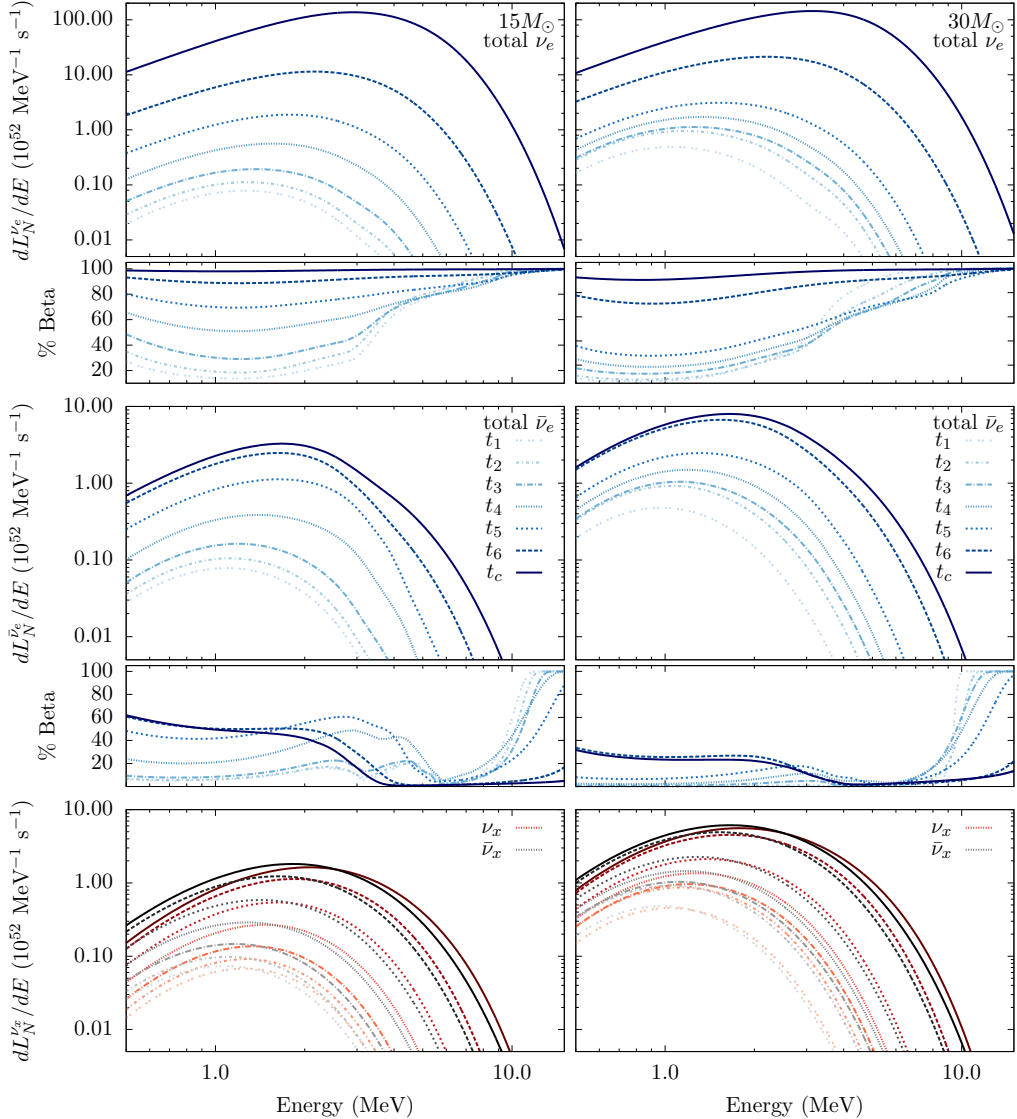}
\caption{Neutrino spectra at selected times pre-collapse for a 15 $\msun$ star (left) and 30 $\msun$ star (right).  Each set of curves shows times $t_{1}$ through $t_{c}$ (lower to upper curves). The exact values of these times are given in Tables \ref{tab:nuclei15} and \ref{tab:nuclei30}. The dashing styles in the legend apply to all panels. The first (third) panel shows the differential luminosity for electron (anti-)neutrinos.  The second (fourth) panel shows the percentage of that luminosity arising from $\beta$ processes.  The bottom panel shows the $\nu_{x}$/$\bar{\nu}_{x}$ luminosity from pair annihilation.}
\label{lumVsEnergy15}
\end{centering}
\end{figure*}

Let us now discuss the \n\ energy spectra and the effect of the \bp\ on them. 
Fig. \ref{lumVsEnergy15} gives the number luminosities, differential in energy, of each \n\ species at seven selected times of the evolution (see Tables \ref{tab:nuclei15} and \ref{tab:nuclei30} for exact values).  Separate panels show the percentages of the $\nue$ and $\barnue$ luminosities that originate from \bp\ alone.

We observe that the $\nue$ and $\barnue$ spectra are smooth at all times, as the integration over the emission volume averages out spectral structures due to \bp\ from individual isotopes, that appear at early times and in certain shells \citep{Patton:2015sqt}.  The spectra have a maximum at $E \sim 1-3$ MeV depending on the time. 
At $E \gta 4$ MeV,  the $\nue$ spectrum is dominated by the \bp\ at all the times of interest (fraction of \bp\ larger than $\sim 60\%$).  At all energies, the \bp\ contribution increases with time, and  it exceeds a 90\% fraction at collapse in the entire energy interval, consistently with fig. \ref{lumVsTime}. 

The percentage of $\barnue$ from \bp\ is lower, overall.  Over time, it increases at low energy ($E \lta 1$ MeV), reaching a $\sim 50\%$ fraction at $E = 1$ MeV at collapse, and decreases at higher energy.  This latter behavior reflects the fact that  the electron degeneracy increases with time, thus reducing the phase space for electrons in the final state due to $\beta^-$ decay.  The lower number density of positrons (relative to electrons)  available for capture also explains the suppression of the \bp\ $\barnue$ flux relative to $\nue$.

\begin{figure*}
\begin{centering}
\includegraphics[width =0.8 \linewidth]{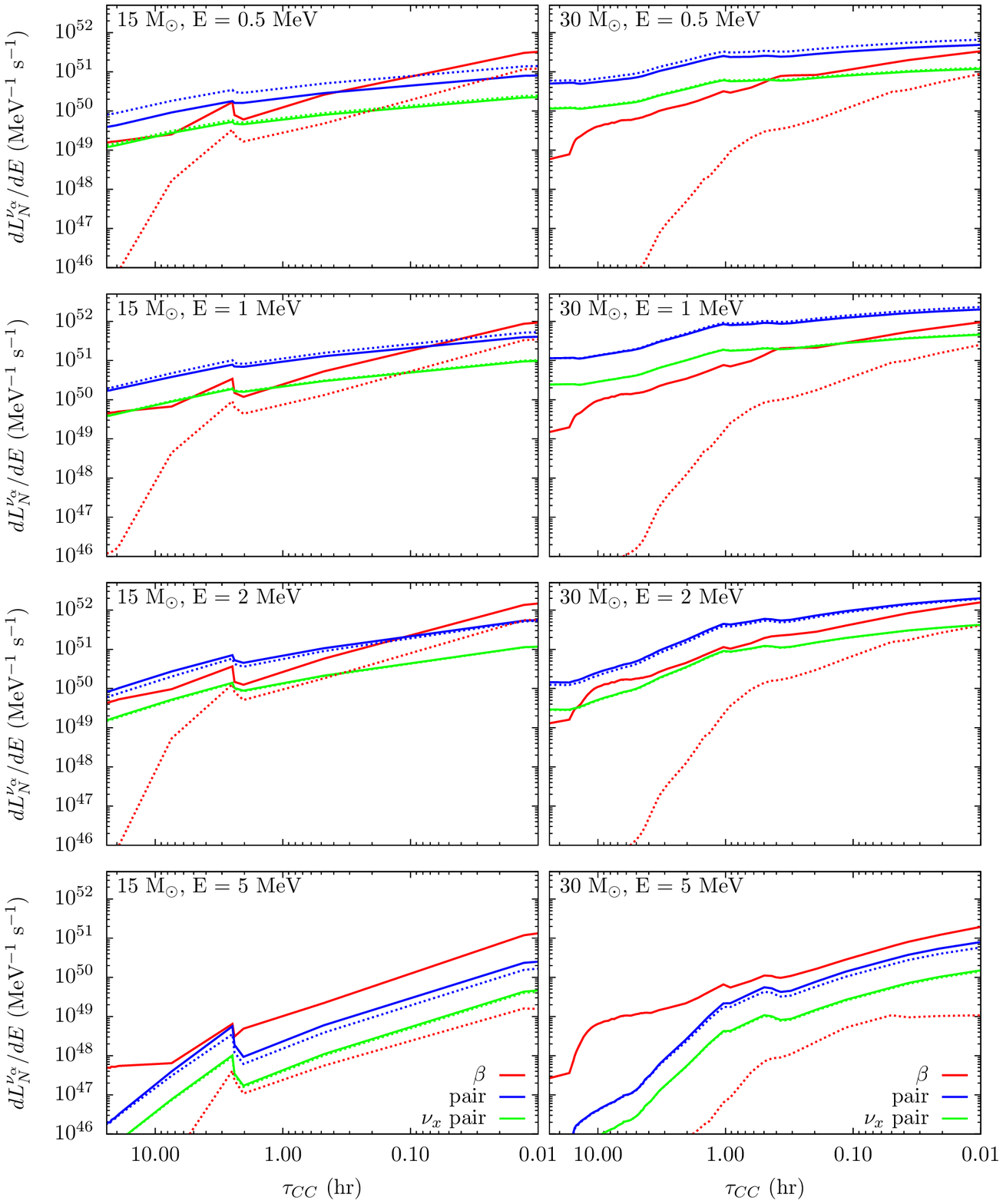}
\caption{The time evolution of the \n\ luminosity for a 15 $\msun$ star (left) and a 30 $\msun$ star (right), differential in energy, at selected energies.  The contributions of the thermal and beta processes are shown separately.  Solid lines represent neutrinos while dashed lines show the antineutrino contributions. }
\label{timeProfile15fig}
\end{centering}
\end{figure*}

A complementary view of these results is given in Fig. \ref{timeProfile15fig}, which shows the time evolution of the \n\ luminosities differential in $E$, at selected values of $E$.   We see that, for the 15 $\msun$ model, the \n\ luminosity from \bp\ has a peak at $\tau_{CC}\approx 2$ hrs, followed by a minimum and a subsequent fast increase. 
This the same feature that appears in the total luminosities for the same progenitor (Fig. \ref{lumVsTime}), and is more pronounced at higher \n\ energy.

What can we learn from \ps\ \ns\ about the isotopic evolution of a star?  To start addressing this question, we investigated what nuclear isotopes contribute the most to the $\nue$ and $\barnue$ fluxes in the detectable region of the spectrum. This is addressed in Tables \ref{tab:nuclei15} and \ref{tab:nuclei30}, where, for selected times in the $\tau_{CC}\lta 2$ hrs, we list the five strongest contributors to both  the total luminosity and the luminosity in the window $E \geq 2$ MeV (where detectors are most sensitive, see Sec. \ref{sub:events}).  The Tables also give the fraction of the \bp\ number luminosity that each isotope produces.  These tables give us a view into how the isotopic makeup of the star evolves over time.  

Let us first describe results for the 15 $\msun$ model.  In it,  silicon shell burning begins at $\tau_{CC}\approx10$ hrs (Sec. \ref{subsec:lum}). Thus in the last two hours before collapse, the isotopic composition is already heavy.  The top five dominant isotopes -- for both $\nue$ and $\barnue$ production -- are those with $A\approx50-60$ such as iron, manganese, cobalt and chromium.  At very late times, $t_{6}$ and $t_{7}$, photodissociation of nuclei becomes efficient, producing free nucleons.  We find that free protons are the strongest contributor to the $\nue$ luminosity at those times.

By summing the contributions listed in Table \ref{tab:nuclei15}, we see that the five dominant isotopes are producing a large percentage of the luminosity: the $\nue$ luminosity from the five dominant isotopes is between $\sim35 - 45\%$ for the total energy range, ending with $\sim50\%$ at $t_{c}$. For $\barnue$, $78\%$ of the luminosity is from the top five isotopes at $t_{1}$.  The percentage gradually decreases to $37\%$ at $t_{c}$. 

The results for the 30 $\msun$ model (Table \ref{tab:nuclei30}), reflect its faster evolution. For this star,  silicon shell burning begins at $\tau\approx 5$ hrs, therefore, it is expected that at $\tau_{CC}\sim 2$ hrs, there might still be a contribution from medium-mass nuclei.  Indeed, the largest contribution to the $\barnue$ luminosity at $t_{1}$ for the 30 $\msun$ star is from $^{28}$Al.  Subsequent times show the same pattern as the 15 $\msun$ model, with mainly isotopes with $A\approx50-60$ dominating.  We see that free protons appear in the top five isotopes at $t_{3} \approx 0.05$ hrs ($\simeq 3$ min) pre-collapse, and are the most dominant contributor from $t_{5}$ on.  Free neutrons also appear in the top-five list for $\barnue$ above $E\geq2$ MeV at $t_{c}$.    For $\nue$, the total contribution of the top-five isotopes is $66\%$ at $t_{1}$, drops to about $40\%$ later, then climbs again to end at $75\%$ at $t_{c}$, of which $\sim 65\%$ is from free protons.  
For $\barnue$, the total fraction is $85\%$ at $t_{1}$, and gradually decreases to $40\%$ at $t_{c}$.  

The fact that, in both models, large portions of the $\nue$ and $\barnue$ luminosities come from a relatively small number of isotopes is promising for future work: it means that efforts to produce more precise \n\ spectra could become more manageable, as they can be  targeted to the subset of isotopes identified in Tables \ref{tab:nuclei15}-\ref{tab:nuclei30}.

\begin{table*}
\begin{tabular}{cc|cccccccccccc}
\hline 
\hline
\multicolumn{14}{c}{15$\msun$}\\
$\tau_{CC}$ (hrs) & & \multicolumn{5}{c}{$ \nue$} & & & \multicolumn{5}{c}{$\barnue$} \\
\hline
\multirow{4}{*}{$t_{1} = 2.038$}  & \multirow{2}{*}{total} & $^{55}$Fe &  $^{56}$Fe &  $^{54}$Fe &  $^{53}$Fe &  $^{55}$Co  & &
& $^{56}$Mn & $^{57}$Mn &  $^{55}$Cr &  $^{52}$V &  $^{53}$V \\
& & 0.141 &  0.0846 &  0.0803 &  0.0778 &  0.0761 & & 
& 0.357 & 0.162 &  0.0937 & 0.0894 &  0.0817 \\ 
& \multirow{2}{*}{E $\geq$ 2 MeV} & $^{53}$Fe & $^{55}$Fe &  $^{55}$Co &  $^{54}$Mn &  $^{57}$Ni & & 
& $^{56}$Mn & $^{52}$V &  $^{57}$Mn &  $^{62}$Co &  $^{55}$Cr \\ 
& & 0.169 & 0.155 &  0.140 & 0.101 &  0.0393  & & 
& 0.423 &  0.107 &  0.0823 &  0.0729 &  0.0689 \\ 
\hline
\multirow{4}{*}{$t_{2} = 1.086$}  & \multirow{2}{*}{total} & $^{55}$Fe &  $^{55}$Co &  $^{56}$Fe &  $^{53}$Fe &  $^{54}$Fe & &
& $^{56}$Mn &  $^{57}$Mn &  $^{55}$Cr &  $^{52}$V &  $^{53}$V \\
& & 0.117 &  0.0860 &  0.0846 &  0.0805 &  0.0779 & & 
& 0.339 &  0.155 &  0.0937 &  0.0894 &  0.0767 \\ 
& \multirow{2}{*}{E $\geq$ 2 MeV} & $^{53}$Fe &  $^{55}$Co &  $^{55}$Fe &  $^{54}$Mn &  $^{57}$Ni & &
& $^{56}$Mn &  $^{62}$Co &  $^{52}$V &  $^{57}$Mn &  $^{58}$Mn \\ 
& & 0.167 & 0.150 &  0.132 &  0.0923 &  0.0482  & & 
& 0.365 &  0.103 &  0.0940 &  0.0909 &  0.0848 \\ 
\hline
\multirow{4}{*}{$t_{3} = 0.4793$}  & \multirow{2}{*}{total} & $^{55}$Fe &  $^{56}$Fe &  $^{55}$Co &  $^{54}$Fe &  $^{53}$Fe  & &
& $^{56}$Mn & $^{57}$Mn &  $^{55}$Cr &  $^{53}$V &  $^{52}$V \\
& & 0.107 &  0.0973 &  0.0641 &  0.0610 &  0.0558 & & 
& 0.247 & 0.158 &  0.101 & 0.0761 &  0.0645 \\ 
& \multirow{2}{*}{E $\geq$ 2 MeV} & $^{55}$Fe &  $^{53}$Fe &  $^{55}$Co &  $^{54}$Mn &  $^{57}$Ni & & 
& $^{56}$Mn & $^{58}$Mn &  $^{57}$Mn &  $^{55}$Cr &  $^{62}$Co \\ 
& & 0.132 & 0.115 &  0.106 &  0.0950 &  0.0424  & & 
& 0.236 &  0.125 &  0.116 &  0.0977 &  0.0957 \\ 
\hline
\multirow{4}{*}{$t_{4} = 0.1022$}  & \multirow{2}{*}{total} & $^{56}$Fe & $^{53}$Cr &  $^{55}$Fe &  $^{57}$Fe &  $^{55}$Mn & & 
& $^{57}$Mn & $^{58}$Mn &  $^{63}$Co &  $^{55}$Cr &  $^{56}$Mn \\
& & 0.110 &  0.0839 &  0.0797 &  0.0554 &  0.0546 & & 
& 0.133 &  0.117 &  0.0981 &  0.0967 &  0.0930 \\ 
& \multirow{2}{*}{E $\geq$ 2 MeV} & $^{55}$Fe &  $^{54}$Mn &  $^{56}$Fe &  $^{53}$Cr &  $^{55}$Co & & 
& $^{58}$Mn & $^{57}$Mn &  $^{63}$Co &  $^{55}$Cr &  $^{62}$Co \\ 
& & 0.104 & 0.0706 &  0.0685 &  0.0646 &  0.0486  & & 
& 0.166 &  0.107 &  0.101 &  0.0935 &  0.0812 \\ 
\hline
\multirow{4}{*}{$t_{5} = 0.01292$}  & \multirow{2}{*}{total} & $^{56}$Fe &  $^{53}$Cr &  $^{57}$Fe &  $^{55}$Fe &  $^{55}$Mn  & &
& $^{58}$Mn & $^{63}$Co &  $^{57}$Mn &  $^{62}$Co &  $^{55}$Cr \\
& & 0.0805 &  0.0783 &  0.0680 &  0.0603 &  0.0526 & & 
& 0.134 & 0.105 &  0.0844 & 0.0760 &  0.0700 \\ 
& \multirow{2}{*}{E $\geq$ 2 MeV} & $^{55}$Fe &  $^{53}$Cr &  $^{54}$Mn &  $^{56}$Fe &  $^{57}$Fe & & 
& $^{58}$Mn & $^{63}$Co &  $^{64}$Co &  $^{62}$Co &  $^{54}$V \\ 
& & 0.0779 & 0.0678 &  0.0597 &  0.0553 &  0.0544  & & 
& 0.170 &  0.0953 &  0.0802 &  0.0695 &  0.0657 \\ 
\hline
\multirow{4}{*}{$t_{6} = 1.632\times10^{-5}$}  & \multirow{2}{*}{total} & $p$ &  $^{53}$Cr &  $^{56}$Fe &  $^{55}$Mn &  $^{51}$V  & &
& $^{58}$Mn & $^{63}$Co &  $^{57}$Mn &  $^{55}$Cr &  $^{54}$V \\
& & 0.212 &  0.0761 &  0.0525 &  0.0497 &  0.0464 & & 
& 0.121 & 0.0889 &  0.0795 & 0.0691 &  0.0583 \\ 
& \multirow{2}{*}{E $\geq$ 2 MeV} & $p$ &  $^{53}$Cr &  $^{51}$V &  $^{55}$Mn &  $^{55}$Fe & & 
& $^{58}$Mn & $^{54}$V &  $^{63}$Co &  $^{55}$Cr &  $^{59}$Mn \\ 
& & 0.233 & 0.0726 &  0.0492 &  0.0459 &  0.0413  & & 
& 0.150 &  0.0869 &  0.0674 &  0.0600 &  0.0584 \\ 
\hline
\multirow{4}{*}{$t_{7} = t_{c}$}  & \multirow{2}{*}{total} & $p$ & $^{53}$Cr &  $^{55}$Mn &  $^{51}$V &  $^{57}$Fe & & 
& $^{58}$Mn &  $^{54}$V &  $^{57}$Mn &  $^{55}$Cr &  $^{56}$Mn \\
& & 0.329 & 0.0565 &  0.0457 &  0.0414 &  0.0398 & & 
& 0.109 &  0.0838 &  0.0660 &  0.0639 &  0.0495 \\ 
& \multirow{2}{*}{E $\geq$ 2 MeV} & $p$ & $^{53}$Cr &  $^{55}$Mn &  $^{51}$V &  $^{56}$Mn & & 
& $^{58}$Mn &  $^{54}$V &  $^{50}$Sc &  $^{59}$Mn &  $^{55}$V \\ 
& & 0.353 & 0.0566 &  0.0441 &  0.0428 &  0.0387  & & 
& 0.123 &  0.113 &  0.0701 &  0.0686 &  0.0619 \\ 
\hline 
\hline
\end{tabular}
\caption{List of the five isotopes that most contribute to the produced $\nue$ ($\barnue$) \ps\ luminosity in the 15 $\msun$ model--  total and at $E\geq2$ MeV -- at selected times. They are listed in order of decreasing $\nue$ ($\barnue$) luminosity.  The number below each isotope is the fraction of the \bp\ number luminosity produced by that isotope. }
\label{tab:nuclei15}
\end{table*}

\begin{table*}
\begin{tabular}{cc|cccccccccccc}
\hline 
\hline
\multicolumn{14}{c}{30$\msun$}\\
$\tau_{CC}$ (hrs) & & \multicolumn{5}{c}{$ \nue$} & & & \multicolumn{5}{c}{$\barnue$} \\
\hline
\multirow{4}{*}{$t_{1} = 2.057$}  & \multirow{2}{*}{total} & $^{54}$Fe &  $^{55}$Fe &  $^{55}$Co &  $^{53}$Fe &  $^{57}$Co  & &
& $^{28}$Al & $^{56}$Mn &  $^{54}$Mn &  $^{24}$Na &  $^{27}$Mg \\
& & 0.219 &  0.192 &  0.110 &  0.0913 &  0.0524 & & 
& 0.603 & 0.0890 &  0.0611 & 0.0557 &  0.0395 \\ 
& \multirow{2}{*}{E $\geq$ 2 MeV} & $^{55}$Fe & $^{53}$Fe &  $^{55}$Co &  $^{54}$Fe &  $^{56}$Co & & 
& $^{28}$Al & $^{24}$Na &  $^{56}$Mn &  $^{60}$Co &  $^{23}$Ne \\ 
& & 0.194 & 0.173 &  0.158 & 0.0798 &  0.0637  & & 
& 0.557 &  0.150 &  0.147 &  0.0532 &  0.0186 \\ 
\hline
\multirow{4}{*}{$t_{2} = 0.4008$}  & \multirow{2}{*}{total} & $^{56}$Ni &  $^{55}$Fe &  $^{55}$Co &  $^{53}$Fe &  $^{54}$Fe & &
& $^{56}$Mn &  $^{57}$Mn &  $^{60}$Co &  $^{61}$Co &  $^{52}$V \\
& & 0.282 &  0.107 &  0.0726 &  0.0629 &  0.0518 & & 
& 0.354 &  0.117 &  0.094 &  0.0597 &  0.0557 \\ 
& \multirow{2}{*}{E $\geq$ 2 MeV} & $^{55}$Fe &  $^{56}$Ni &  $^{55}$Co &  $^{53}$Fe &  $^{52}$Fe & &
& $^{56}$Mn &  $^{57}$Mn &  $^{60}$Co &  $^{61}$Co &  $^{55}$Cr \\ 
& & 0.138 & 0.125 &  0.114 &  0.109 &  0.0606  & & 
& 0.383 &  0.119 &  0.0865 &  0.0623 &  0.0613 \\ 
\hline
\multirow{4}{*}{$t_{3} = 0.05069$}  & \multirow{2}{*}{total} & $^{55}$Fe &  $^{56}$Ni &  $^{56}$Fe &  $p$ &  $^{55}$Co  & &
& $^{56}$Mn & $^{57}$Mn &  $^{62}$Co &  $^{55}$Cr &  $^{58}$Mn \\
& & 0.101 &  0.101 &  0.0782 &  0.0678 &  0.0472 & & 
& 0.229 & 0.126 &  0.0889 & 0.0799 &  0.0688 \\ 
& \multirow{2}{*}{E $\geq$ 2 MeV} & $^{55}$Fe &  $^{54}$Mn &  $^{55}$Co &  $p$ &  $^{56}$Fe & & 
& $^{56}$Mn & $^{57}$Mn &  $^{62}$Co &  $^{58}$Mn &  $^{55}$Cr \\ 
& & 0.128 & 0.0736 &  0.0621 &  0.0576 &  0.0540  & & 
& 0.207 &  0.119 &  0.109 &  0.0995 &  0.0883 \\ 
\hline
\multirow{4}{*}{$t_{4} = 0.04892$}  & \multirow{2}{*}{total} & $^{55}$Fe & $^{56}$Ni &  $^{56}$Fe &  $p$ &  $^{55}$Co & & 
& $^{56}$Mn & $^{57}$Mn &  $^{62}$Co &  $^{55}$Cr &  $^{58}$Mn \\
& & 0.101 &  0.101 &  0.0779 &  0.0698 &  0.0471 & & 
& 0.228 &  0.126 &  0.0870 &  0.0801 &  0.0683 \\ 
& \multirow{2}{*}{E $\geq$ 2 MeV} & $^{55}$Fe &  $^{54}$Mn &  $^{55}$Co &  $p$ &  $^{56}$Fe & & 
& $^{56}$Mn & $^{57}$Mn &  $^{62}$Co &  $^{58}$Mn &  $^{55}$Cr \\ 
& & 0.128 & 0.0736 &  0.0619 &  0.0595 &  0.0539  & & 
& 0.207 &  0.120 &  0.107 &  0.0990 &  0.0887 \\ 
\hline
\multirow{4}{*}{$t_{5} = 0.00214$}  & \multirow{2}{*}{total} & $p$ &  $^{55}$Fe &  $^{56}$Fe &  $^{56}$Ni &  $^{54}$Mn  & &
& $^{58}$Mn & $^{57}$Mn &  $^{56}$Mn &  $^{62}$Co &  $^{55}$Cr \\
& & 0.284 &  0.0646 &  0.0532 &  0.0414 &  0.0363 & & 
& 0.116 & 0.109 &  0.107 & 0.0836 &  0.0829 \\ 
& \multirow{2}{*}{E $\geq$ 2 MeV} & $p$ &  $^{55}$Fe &  $^{54}$Mn &  $^{56}$Fe &  $^{57}$Co & & 
& $^{58}$Mn & $^{57}$Mn &  $^{55}$Cr &  $^{62}$Co &  $^{56}$Mn \\ 
& & 0.494 & 0.0442 &  0.0371 &  0.0281 &  0.0261  & & 
& 0.163 &  0.0917 &  0.0867 &  0.0794 &  0.0712 \\ 
\hline
\multirow{4}{*}{$t_{6} = 0.0001142$}  & \multirow{2}{*}{total} & $p$ &  $^{55}$Fe &  $^{56}$Fe &  $^{53}$Cr &  $^{54}$Mn  & &
& $^{58}$Mn & $^{57}$Mn &  $^{55}$Cr &  $^{56}$Mn &  $^{63}$Cr \\
& & 0.487 &  0.0393 &  0.0371 &  0.0293 &  0.0283 & & 
& 0.121 & 0.0996 &  0.0809 & 0.0774 &  0.0645 \\ 
& \multirow{2}{*}{E $\geq$ 2 MeV} & $p$ &  $^{55}$Fe &  $^{54}$Mn &  $^{56}$Fe &  $^{53}$Cr & & 
& $^{58}$Mn & $^{54}$V &  $^{55}$Cr &  $^{57}$Mn &  $^{63}$Co \\ 
& & 0.494 & 0.0442 &  0.0371 &  0.0281 &  0.0261  & & 
& 0.167 &  0.0820 &  0.0811 &  0.0747 &  0.0576 \\ 
\hline
\multirow{4}{*}{$t_{7} = t_{c}$}  & \multirow{2}{*}{total} & $p$ & $^{53}$Cr &  $^{55}$Mn &  $^{56}$Fe &  $^{54}$Mn & & 
& $^{58}$Mn &  $^{57}$Mn &  $^{55}$Cr &  $^{56}$Mn &  $^{53}$V \\
& & 0.639 & 0.0286 &  0.0252 &  0.0234 &  0.0213 & & 
& 0.0963 &  0.0943 &  0.0819 &  0.0747 &  0.0555 \\ 
& \multirow{2}{*}{E $\geq$ 2 MeV} & $p$ & $^{53}$Cr &  $^{55}$Mn &  $^{54}$Mn &  $^{51}$V & & 
& $^{58}$Mn &  $^{55}$Cr &  $^{54}$V &  $^{57}$Mn &  $n$ \\ 
& & 0.659 & 0.0273 &  0.0236 &  0.0236 &  0.0212  & & 
& 0.124 &  0.0803 &  0.0760 &  0.0705 &  0.0638 \\ 
\hline 
\hline
\end{tabular}
\caption{Same as Table \ref{tab:nuclei15} but for the 30 $\msun$ model.}
\label{tab:nuclei30}
\end{table*}

\section{Propagation and detectability }
\label{sec:atearth}

\subsection{Oscillations of presupernova neutrinos }
\label{sub:oscill}

The flavor composition of the \ps\ \n\ flux at Earth differs from the one at production, due to flavor conversion (\oss).  In terms of the original, unoscillated flavor luminosities, $F^0_\alpha = dL^{\nu_\alpha}_{N}/dE$ ($\alpha = e, \bar e, x$), the fluxes of each \n\ species at Earth can be written as  
\be 
F_e = p F^0_e + (1-p) F^0_x ~, \hskip1truecm  2 F_x = (1-p) F^0_e + (1+p) F^0_x~,
\label{fluxearth}
\ee
where $F_x$ is defined so that  the total flux is $F_e + 2 F_x = F^0_e + 2 F^0_x$, and the geometric factor $(4 \pi D^2)^{-1}$, due to the distance $D$ to the star,  is omitted for brevity.  An expression analogous to eq. (\ref{fluxearth}) holds for antineutrinos, with the notation replacements $e \rightarrow \bar e$ and $p \rightarrow \bar p$.   The quantities $p$ and $\bar p$ are the $\nue$ and $\barnue$ survival probabilities. They have 
 been studied extensively for a supernova \n\ burst (see, e.g. \citep{Duan:2009cd} and references therein), and at a basic level for \ps\  \ns\ \citep{Asakura:2015bga, Kato:2015faa, Kato:2017ehj}.  Similarly to the burst \ns, \ps\ \ns\ undergo adiabatic, matter-driven, conversion inside the star.  The probabilities $p$ and $\bar p$  are are independent of energy and of time. They are given by the elements of the \n\ mixing matrix, $U_{\alpha i}$, in a way that depends on the (still unknown) \n\ mass hierarchy; given the masses $m_i$ ($i=1,2,3$), the standard convention defines the normal hierarchy (NH) as $m_3 > m_2$ and vice versa for the inverted hierarchy (IH).  For each possibility, we have (see e.g., \citep{Lunardini:2003eh, Kato:2017ehj}): 
\be
p = \begin{cases}
  |U_{e3}|^2 \simeq 0.02  & \text{ NH }  \\
   |U_{e2}|^2 \simeq 0.30  & \text{ IH }~
\end{cases}  \hskip0.8truecm \bar p= \begin{cases}
  |U_{e1}|^2 \simeq  0.68  & \text{ NH }  \\
   |U_{e3}|^2 \simeq 0.02  & \text{ IH }~.
\end{cases} \label{psurv}
\ee

For simplicity here we do not consider other oscillation effects, namely collective oscillations inside the star and oscillations in the matter of the Earth. The former are expected to be negligible due to the relatively low \ps\ \n\ luminosity (compared to the \sn\ burst), and the latter are suppressed (a $\sim 1\%$ effect or less) at the energies of interest here (see e.g., \citep{Wan:2016nhe}).  
  
Eq. (\ref{psurv}) shows that for the NH the $\nue$ flux at Earth receives only a  very suppressed contribution from the original $\nue$. The suppression is weaker for the IH, and therefore -- considering that $F^0_x \ll F^0_e$--   the flux $F_e$ should be much larger in this case.  For the $\barnue$ flux, a smaller difference between NH and IH is expected, due to $F^0_x$ and $F^0_{\bar e}$ being comparable (Fig. \ref{lumVsTime}). 

\subsection{Window of observability}
\label{sub:detection}

A detailed discussion of  the detectability of \ps\ \ns\ is beyond the scope of this paper, and is deferred to future work. Here general considerations are given on the region, in the time and energy domain, where detection might be possible -- depending on the distance to the star -- and the numbers of events  expected in \n\ detectors are given.  
  
One can define a conceptual window of observability (WO)  as the interval of time and energy where the \ps\ flux  exceeds all the \n\ fluxes of other origin that are (i) present in a detector at all times, and (ii) indistinguishable from the signal. These fluxes are guaranteed backgrounds, regardless of the details of the detector in use; to them, detector-specific backgrounds will have to be added. Therefore the WO defined here represent a most optimistic, ideal situation.  
  
Because observations at \n\ detectors are generally dominated by either $\nue$ or $\barnue$, let us discuss the WOs for these two species.  In the case of $\nue$, the largest competing flux is due to solar \ns\ \citep{Bahcall:2005qb}. 
For $\barnue$, we consider fluxes from nuclear reactors   and from the Earth's natural radioactivity (geo\ns) \citep{Fiorentini:2007ix}.  For both $\nue$ and $\barnue$, other background fluxes are from  atmospheric \ns\ and from the diffuse supernova neutrino background (DSNB, due to all the \sn\ \n\ bursts in the universe).  At the times and energies of interest, however, these are much lower than the solar, reactor, geo\ns\ and presupernova fluxes, and therefore they will be neglected from here on.

 The reactor neutrino and geo\n\ spectra depend on the location of the detector in relation to working reactors and local geography.  
The reactor spectrum we use was calculated for the 
Pyh\"asalmi mine in Finland \citep{Mollenberg:2015yg, Wurm:2009}, and includes oscillations.  The geoneutrino spectrum is generic, and includes vacuum oscillations only, with survival probability at JUNO $\bar p\simeq 0.55$ for both NH and IH \citep{Wan:2016nhe} \footnote{Effects from MSW oscillation are shown to be at the level of 0.3\% \citep{Wan:2016nhe}, and therefore can be neglected.}

\begin{figure*}
\begin{centering}
\includegraphics[width =0.8 \linewidth]{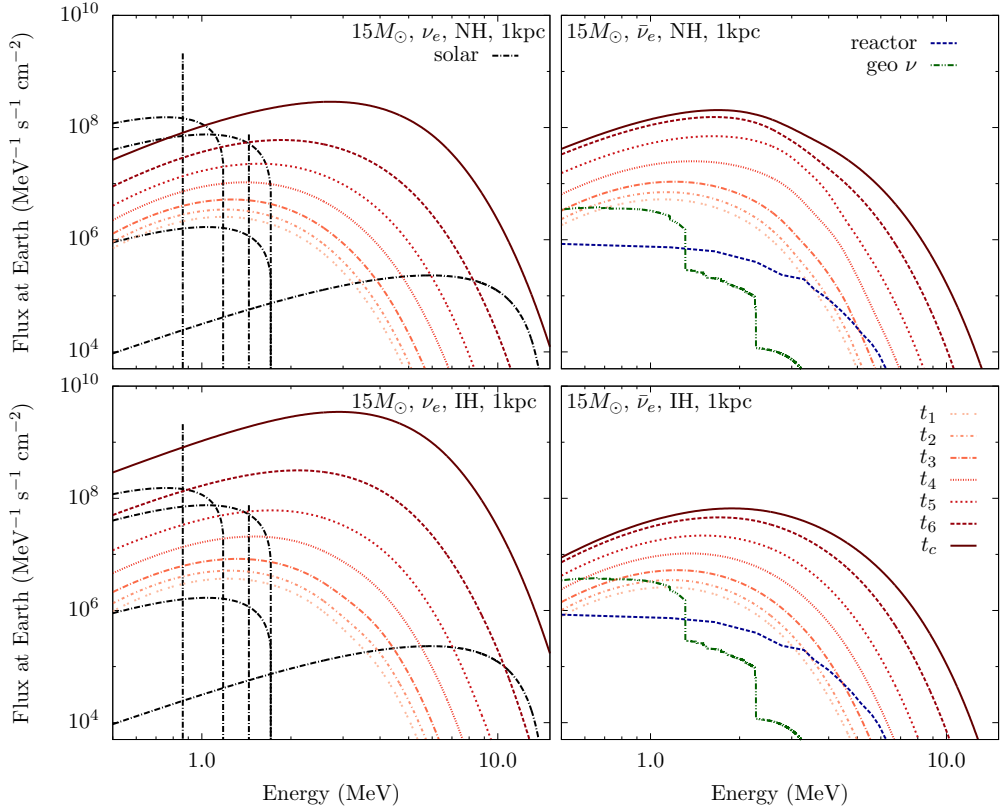}
\caption{The  flux of $\nue$ (left panels) and $\barnue$ (right panels) expected at Earth from a $15 \msun$ star at distance $D=1$ kpc,  calculated at times $t_{1}$ through $t_{c}$ (lower to upper curves).  Shown are the cases of normal  and inverted mass hierarchy (upper and lower rows respectively).    Competing \n\ fluxes from other sources are shown (see legend). Oscillations are included in all cases.
 }
\label{oscillated15}
\end{centering}
\end{figure*}

\begin{figure*}
\begin{centering}
\includegraphics[width =0.8 \linewidth]{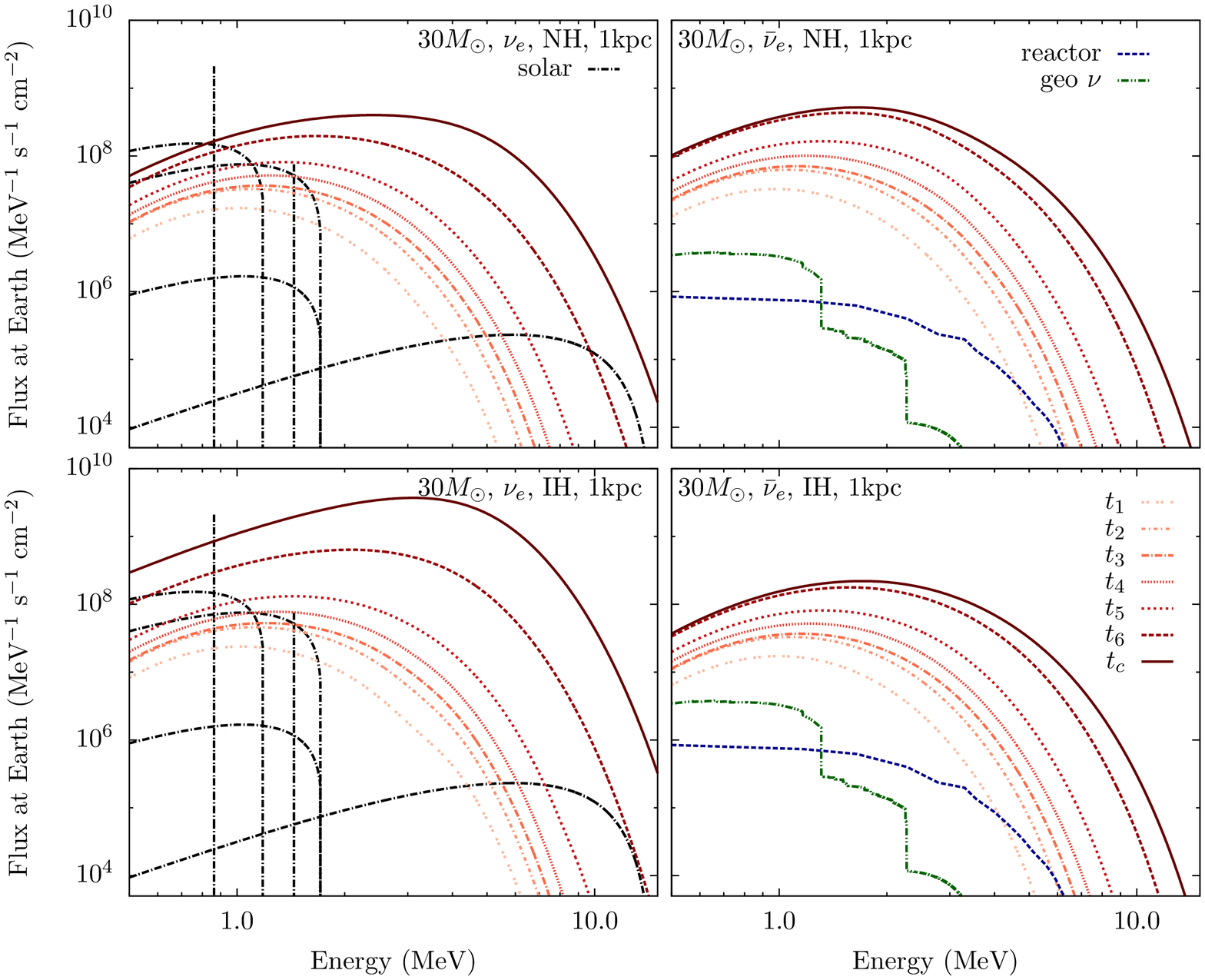}
\caption{Same as Fig. \ref{oscillated15} for a 30 $\msun$ star. }
\label{oscillated30}
\end{centering}
\end{figure*}

Figs. \ref{oscillated15} and \ref{oscillated30} shows the \ps\ \n\ signal at Earth for a star at $D=1$ kpc. 
It appears that, already two hours before collapse, the \ps\ $\nue$ flux emerges above solar \ns. The WO becomes wider in energy as the \ps\ flux increases with time.   An approximate WO is $\tau_{CC} \sim 0 - 2 $ hrs, and $E \sim 1 - 8$ MeV, and it is larger for the IH and for the more massive progenitor, where the \ps\ flux is higher.  We note that it may be possible to distinguish and subtract solar \ns\ effectively using their arrival direction, e.g., in \n-electron scattering events in water Cherenkov detectors \citep{SuperK:2011}.  With a $\sim 10^4$ reduction in the solar background, the $\nue$ WO would extend in energy and time, $\tau_{CC} \sim 0 - 24 $ hrs, and $E \sim 0.5 - 10$ MeV. 

For the same distance $D=1$ kpc, the WO for $\barnue$ is similar to that of $\nue$, but it is overall wider in energy, as the \ps\ flux eventually exceeds the geo\n\ one at sub-MeV energy. Approximately, the WO is $\tau_{CC} \sim 0 - 2 $ hrs, and $E \sim 0.5 - 20$ MeV. 

By increasing the distance $D$, the WO becomes narrower; unless the background fluxes in Figs. \ref{oscillated15} and \ref{oscillated30} are subtracted,  it eventually closes completely for $D \sim 30$ kpc.  This maximum distance -- which is of the order of the size our galaxy --  is independent of the specific detector considered. We will see below that the actual horizon for observation is smaller for realistic detector masses. 

It is possible that the next supernova in our galaxy will be closer than 1 kpc, thus offering better chances of \ps\ \n\ observation.  A prime example is the red supergiant Betelgeuse ($\alpha$ Orionis). Betelgeuse has the largest angular diameter on the sky of any star apart from the Sun, and is the ninth-brightest star in the night sky. As such, it has been well studied.  Betelgeuse is estimated to have a mass of 11 - 20 M$_{\odot}$ \citep{2001ApJ...558..815L, 2011ASPC..451..117N, 2016ApJ...819....7D, 2016ApJ...830..103N} ; it lies at a distance of 222$^{+48}_{-34}$ pc \citep{2008AJ....135.1430H, 2017AJ....154...11H}, and has an age of 8 - 10 Myr, with $<$ 1 Myr of life left until core collapse \citep{2016ApJ...819....7D, 2017AJ....154...11H}. We find that for $D=200$ pc a \ps\ \n\ signal would be practically background-free -- in energy windows that are realistic for detection -- for several hours, and the WO can extend up to $\sim 10$ hours.

\subsection{Numbers of events, horizon}
\label{sub:events}

Let us now briefly discuss expected numbers of events at current and near future detectors of  ${\mathcal O}(10)$ kt scale or higher.  We consider the three main detection technologies: liquid scintillator (JUNO \citep{An:2015jdp}), water Cherenkov (Super Kamiokande \citep{ABE2014253}) and liquid argon (DUNE \citep{cdr}).  For each, we consider the dominant detection channel -- that will account for the majority of the events in the detector -- and the first subdominant process that is sensitive to $\nue$. 
The latter will be especially sensitive to $\nue$ from the \bp.  

For water Cherenkov  and liquid scintillator, the dominant detection process is inverse beta decay (IBD), $\barnue + p \rightarrow n + e^+$, which bears some sensitivity to $\barnue$ from \bp.  The sensitivity to $\nue$ from the \bp\ is in the subdominant channel,  neutrino-electron elastic scattering (ES), where the contribution of $\nue$ is enhanced  (compared to $\nux$) by the larger cross section.   Note that the two channels, IBD and ES, can be distinguished in the detector, at least in part, due to their different final state signatures: neutron capture in coincidence for IBD, and the peaked angular distribution for ES (see, e.g., \citep{Beacom:1998fj, Ando:2001zi}).  In Super Kamiokande, efficient neutron capture will be possible in the upcoming upgrade with Gadolinium addition \citep{Beacom:2003nk}.  

In liquid scintillator (LS), the main detection processes are the same as in water, with the differences that LS offers little directional sensitivity, but has the advantage of a lower, sub-MeV energy threshold, which can capture most of the \ps\ spectrum. 

In liquid Argon (LAr), the dominant process is $\nue$ Charged Current scattering on the Argon nucleus. Therefore, LAr is, in principle,  extremely sensitive to \ns\ from the \bp. However, the relatively high energy threshold ($E_{th}\sim 5 $ MeV \citep{Acciarri:2015uup}) is a considerable disadvantage compared to LS. 

Table  \ref{tab:rates15} and \ref{tab:rates30}  summarize our results for the number of \ps\ \n\ events expected above realistic thresholds during the last two hours precollapse. 
The numbers of background events are not given, because they are affected by large uncertainties on the contributions of detector-specific backgrounds. These ultimately depend on type of search performed, and have not been studied in detail yet for a \ps\ signal \footnote{Most background rejection studies have been performed for type of signals that are either constant in time or very short (e.g., a \sn\ burst). A \ps\ signal is intermediate, rising steadily over a time scale of hours. This feature might require developing different approaches to cut backgrounds. }.

The tables confirm that a  large liquid scintillator like JUNO has the best potential, due to its sensitivity at low energy, with $N \sim 10 - 70$ events (depending on the type of progenitor) recorded from a star at $D=1$ kpc.  To these events, the contribution of the \bp\  is at the level of $10-30\%$, and is larger for the inverted mass hierarchy, for which the $\nue$
 flux is larger, see Sec. \ref{sub:oscill}. For Betelgeuse, a spectacular signal of more than 200 events in two hours could be seen. 
One can define (optimistically) the horizon of the detector, $D_h$, as the distance for which one signal event is expected. We find that JUNO has a horizon $D_h \sim 2-8$ kpc.  

Although disadvantaged by the higher energy threshold, SuperKamiokande and DUNE can observe \ps\ \ns\ for the closest stars.  For the most massive progenitor, SuperKamiokande could reach a horizon $D_h \sim 1$ kpc; and record $N  \sim 5-60$ events for $D=0.2$ pc. Of these, $\sim 10-20\%$ would be from \bp.  Looking  farther in the future, the larger water Cherenkov detector HyperKamiokande \citep{Abe:2011ts}  -- with mass 20 times the mass of SuperKamiokande --  might become a reality.  Assuming an identical performance as SuperKamiokande,  HyperKamiokande will have a statistics of up to thousands of events, and a horizon of $\sim 4-5$ kpc  \footnote{Due to its mass, HyperKamiokande will have a $\sim$20 times higher level of background than SuperKamiokande, and, probably, a higher energy threshold. Therefore, its performance will be worse, and the figures given here have to be taken as best case scenarios.  }.  

At DUNE, a detection is possible only for the closest stars; the number of events varies between $N  \sim 1$ and $ N \sim 30$, depending on the parameters, for $D=0.2$ kpc. For the most optimistic scenario (the more massive progenitor and the inverted mass hierarchy), the horizon can reach $D_{h} \sim 1$ kpc.  DUNE will observe a strong component due to \bp\ , at the level of  $\sim 40-80\%$ of the total signal.  Therefore, in principle LAr has the best capability to probe the isotopic evolution of \sn\ progenitors.

\begin{table*}[htp]
\begin{center}
\begin{tabular}{|c|c|c|c|c|c|c|c|c|c|c|c| 
}
\hline
\hline
detector & composition &  mass  &   interval  &   $N^{CC}_{\beta}$ & $N^{el}_{\beta}$ &  &   $N^{CC}$  & $N^{el}$  & & $N^{tot}=N^{CC}+N^{el}$ \\ 
\hline
\hline
JUNO & $C_nH_{2n}$ &  17 kt  & $E_e \geq 0.5$ MeV &  3.19 & 2.34 && 10.1 & 7.19 && 17.3  \\
								 &  &   &  &   [0.09 ]& [4.32]  && [2.592] & [10.2] && [12.8] \\
\hline
SuperKamiokande & $H_2 O$ & 22.5 kt & $E_e \geq 4.5$ MeV  & 0.04 & 0.02 && 0.43 & 0.03 && 0.45  \\
  &  &   &  & [ 0.00]  & [0.05] && [0.15]  & [0.06] && [0.21] \\
  \hline
\hline
DUNE & LAr  & 40 kt  & $E \geq 5$ MeV &  0.017 & 0.013 && 0.046 & 0.018 && 0.063  \\
 				 &  &   &  & [0.27]  & [0.032] && [0.33] & [0.039] && [0.37]  \\ 

     \hline
\hline
\end{tabular}
\end{center}
\caption{  Numbers of events expected in the two hours prior to collapse, for a \ps\ with progenitor mass $M=15\msun$, at distance $D=1$ kpc  and the normal mass hierarchy.  The numbers in brackets refer to the inverted mass hierarchy. Different columns give the numbers for different detection channels:  the superscripts $CC$ and $el$ refer respectively to the dominant charged current process (inverse beta decay or $\nue$ absorption on the Ar nucleus) and to \n-electron scattering. The subscript $\beta$ indicates the contribution of the $\beta$ processes to those two channels.  The total number of events is given in the last column.   The results for Betelgeuse ($D=0.2$ kpc) can be obtained by rescaling by a factor of 25.  }
\label{tab:rates15} 
\end{table*}%

\begin{table*}[htp]
\begin{center}
\begin{tabular}{|c|c|c|c|c|c|c|c|c|c|c|c| 
}
\hline
\hline
detector & composition &  mass  &   interval  &   $N^{CC}_{\beta}$ & $N^{el}_{\beta}$ &  &   $N^{CC}$  & $N^{el}$  & & $N^{tot}=N^{CC}+N^{el}$ \\ 
\hline
\hline

JUNO & $C_nH_{2n}$ &  17 kt  & $E_e \geq 0.5$ MeV & 1.83 & 4.40 && 40.1 & 32.1 && 72.3  \\ 
								 &  &   &  &  [0.05] & [9.47] && [13.1] & [42.7] && [55.9] \\
\hline
SuperKamiokande & $H_2 O$ & 22.5 kt & $E_e \geq 4.5$ MeV  & 0.063 & 0.053 && 2.27 & 0.098 && 2.37 \\ 
  &  &   &  & [0.00] & [0.13] && [0.78] & [0.20] && [0.98] \\  
\hline
\hline
DUNE & LAr  & 40 kt  & $E \geq 5$ MeV & 0.05  & 0.04 && 0.19 &  0.06 && 0.25  \\ 
 				 &  &   &  &  [0.76] &  [0.09]  & & [1.1]  &  [0.13]  &  &  [1.2]  \\ 

     \hline
\hline
\end{tabular}
\end{center}
\caption{ Same as Tab. \ref{tab:rates15}, for the $M=30\msun$ progenitor.  }
\label{tab:rates30} 
\end{table*}%

\section{Discussion}
\label{sec:discussion}

We have presented a new calculation of the total \n\ flux from beta processes in a \ps\ star, inclusive of time-dependent emissivities and \n\ energy spectra. This is part of a complete and detailed calculation of \ps\ \n\  fluxes from most relevant processes -- beta and thermal -- done using the state of the art stellar evolution code MESA.  

The beta \n\ flux is strongest in the $\nue$ channel, where it is comparable to the flux from thermal processes in the few hours pre-collapse, and it even exceeds it in the high energy tail of the spectrum, $E\gta 3$ MeV.  This very relevant for current and near future detectors, which are most sensitive above the MeV scale. 

Among the realistic detection technologies, liquid scintillator is best suited to detect \ps\ \ns. This is due to its lower energy threshold, which allows to capture the bulk of the flux hours or minutes before collapse.  In such detector \ns\ from beta processes  would contribute up to $\sim 30\%$ of the total number of events, for a threshold of $\sim 0.5$ MeV.  The horizon for detection (i.e., the distance from the star where a few events are expected in the detector) is of a few kpc for a 17 kt detector, with tens of events expected for $D\simeq 1$ kpc.  
The number of event increases strongly with the mass of the progenitor star; therefore, for 
medium-high statistics and known $D$, the \ps\ \n\ signal will contribute to establishing the type of progenitor. For high statistics, the time profile of the \ps\ signal could provide additional information, e.g., on the time of ignition of the different fuels (fig. \ref{lumVsTime}). 

At water Cherenkov and liquid Argon detectors of realistic sizes and thresholds ($E_{th} \sim 4-5$ MeV), the horizon is generally limited to the closest stars, $D \sim {\mathcal O}(0.1)$ kpc, but could reach 1 kpc for the most massive progenitors and the inverted \n\ mass hierarchy. For liquid Argon, the contribution of the $\beta$ \ns\ is strong, and could even dominate the signal.  Therefore - at least in principle - liquid Argon detectors offer the possibility of probing the complex nuclear processes in stellar cores. 

If the high energy tail of a \ps\ flux is detected, what nuclei and what processes exactly can we probe?  To answer this question, we have identified the isotopes that mostly contribute to the \ps\ $\nue$ flux in the detectable energy window, generally iron, manganese and cobalt isotopes as well as free protons and neutrons. The possibility that \n\ detectors may test the physics of these isotopes is completely novel. 

In closing, we stress that our calculation used the best available instruments: a state of the art stellar evolution code, combined with the most up-to-date studies of nuclear rates and beta spectra. Still, these instruments are affected by uncertainties, which, naturally, affect the results in this paper. In particular, while total emissivities are relatively robust, it is likely that the highest energy tails of the \n\ spectrum, in the detectable window, are very sensitive to the details of the calculation, i.e.,  the temperature profile of the star, the nuclear abundances and the quantities in the nuclear tables we have used.  Specifically for \n\ spectra, a source of error lies in the single-strength approximation that is adopted here for \bp\ (sec. \ref{sec:theory}).   A  recent paper \citep{Misch:2016iwm}, presents an exploratory study of this error and concludes that while the single effective $Q$-value approach results in the correct emissivity and average energy, the specific energy spectrum could miss important features.  A systematic extension of this result to the many isotopes included in MESA would be highly desirable to improve our results.  Another interesting addition to the code would be the contribution of \n\ pair production via neutral current de-excitation \citep{Misch:2016iwm}, which is currently omitted in MESA.  This de-excitation results in higher energy neutrino spectra than the processes described in this work, and thus makes detections more likely.  

Until these important improvements become available, our results have to be interpreted conservatively, as a proof of the possibility that current and near future detectors might be able to observe \ps\ \ns, and therefore offer the first, \emph{ direct} test of the isotopic evolution of a star in the advanced stages of nuclear burning.

\begin{acknowledgments}
We thank K. Zuber and Wendell Misch  for fruitful
discussion. We also acknowledge the National Science Foundation grant
number PHY-1205745, and the Department of Energy award DE-SC0015406.
This project was also supported by NASA under the Theoretical and
Computational Astrophysics Networks (TCAN) grant NNX14AB53G, by NSF
under the Software Infrastructure for Sustained Innovation (SI2) grant
1339600, and grant PHY 08-022648 for the Physics Frontier Center
"Joint Institute for Nuclear Astrophysics - Center for the Evolution
of the ElementsÓ (JINA-CEE).  This research was also partially
undertaken at the Kavli Institute for Theoretical Physics which is
supported in part by the National Science Foundation under Grant
No. NSF PHY-1125915. 

\end{acknowledgments}

\bibliography{bibINSPIRE}

\end{document}